\documentclass[twocolumn,pre]{revtex4}
\usepackage{graphicx}
\usepackage{epsfig}
\usepackage{latexsym}
\usepackage{amssymb}
\usepackage{bm}
\usepackage{float}
\usepackage{subfigure}
\usepackage{amsmath}
\usepackage{amsfonts}
\begin{document}

\title{Correlations and critical behavior of the q-model}
\author{Alexander St. John$^{(1)}$}
\author{Harsh Mathur$^{(2)}$}
\affiliation{$^{(1)}$ Department of Physics, University of Wisconsin Parkside, 900 Wood Rd, Kenosha, WI 53144}
\affiliation{$^{(2)}$Department of Physics, Case Western Reserve University, 10900 Euclid Avenue, Cleveland OH 44106}

\begin{abstract}

The q-model is a random walk model used to describe the flow of stress in a stationary granular medium.
Here we derive the exact horizontal and vertical correlation functions for the q-model in two dimensions. We 
show that close to a critical point identified in earlier work these correlation functions have a universal 
scaling form reminiscent of thermodynamic critical phenomena. We determine the form of the universal scaling 
function and the associated critical exponents $\nu$ and $z$. 

\end{abstract}

\maketitle

\section{Introduction}
\label{sec:introduction}

The q-model was introduced by Coppersmith {\em et al.} \cite{coppersmith}
to describe the flow of stress in a static granular medium.
Although it does not provide a complete solution to the problem of stress flow in a granular medium, 
it provides a good first approximation.
Moreover, the q-model is closely related to models that describe the process of aggregation in statistical mechanics
\cite{takayasu}, the transport of electrons on the surface of an integer quantum Hall multilayer
\cite{chalker}, passive scalar turbulence \cite{siggia},
and the branching of river networks \cite{dodds}, among others. In earlier work by Lewandowska {\em et al.} \cite{marta}
it was shown that although the q-model describes physics far from equilibrium, nonetheless, 
for a particular value of its parameters, the q-model behaves in a manner
reminiscent of the critical point in an equilibrium thermodynamic phase transition. It was found that the q-model
is remarkably soluble and a number of exact critical exponents and universal scaling functions near to the 
critical point were obtained. The purpose of the present work is to strengthen the analogy to thermodynamic
critical phenomena by calculating the two point correlation functions for the q-model.

In experiments that the q-model is used to describe a pack of beads is loaded from above with a uniform stress.
The distribution of load at the bottom of the bead pack as well as the propagation of stress through the pack have
been measured experimentally \cite{experiments}. 
In the q-model it is assumed that the beads lie at the sites of a regular lattice.
The beads in each layer are supported by their nearest neighbors in the layer below. For simplicity let us consider
the q-model in two spatial dimensions. Let us suppose that the beads are arranged in a square lattice as
shown in the figure. For each bead a fraction $f$ of its load is supported by its neighbor to the left in the layer below
and a fraction $1-f$ by its neighbor to the right. These fractions are assumed to vary randomly from bead to bead.
We denote the depth of the layer by $t$ and the position of a bead within a layer by $n$. The load on bead $n$ in
layer $t+1$ is then determined by the recursion relation
\begin{equation}
w_n (t+1) = f_{n+1,t} w_{n+1} (t) + (1 - f_{n-1, t}) w_{n-1, t}
\label{eq:weightrecursion}
\end{equation}
Given the random fractions $f$ one can use eq (\ref{eq:weightrecursion}) to propagate an applied uniform 
load in the top layer downwards. The goal is to obtain a statistical description of the load as a function of depth
by analytic or numerical solution of eq (\ref{eq:weightrecursion}). 

\begin{figure}[h]
\begin{center}
\includegraphics[width=0.45\textwidth]{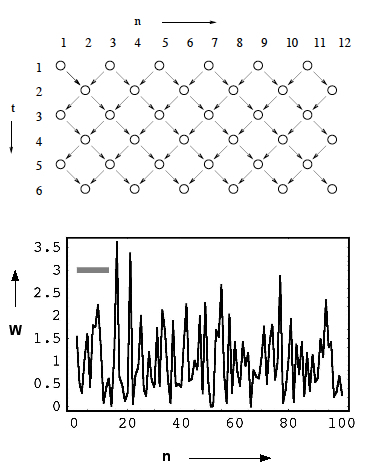}
\end{center}
\caption{The q-model in two dimensions (reproduced from ref \cite{marta}). 
The beads are assumed to be arranged in a square lattice. Each bead
is assumed to be supported by its nearest neighbors in the layer below. A random fraction $f$ of the load is transferred
to the neighbor at left; the remainder, $(1-f)$ to the neighbor to the right. The fractions vary from bead to bead and are
assumed to be independent identically distributed random variables. $t$ specifies the horizontal layer occupied
by a bead. The horizontal layers are numbered by consecutive integers. The position of a bead within a 
horizontal layer is given by $n$. $n$ is even in a layer with even $t$; odd, for odd $t$. Thus the horizontal 
separation of two beads in the same layer must be an even integer. Similarly the horizontal separation for
two beads in layers with an even vertical separation must also be an even integer.}
\label{fig:qmodel}
\end{figure}

In the q-model it is generally assumed that the fractions are independent
identically distributed random variables with a distribution $P(f)$ that satisfies the symmetry requirement
$P(f) = P(1-f)$. The simplest choice is to assume that each fraction is either zero or one with equal probability;
this is called the singular distribution and it was shown in ref \cite{marta} that the behavior of the q-model
for this distribution is reminiscent of a critical point in equilibrium thermodynamics. Other distributions can be
characterized by a parameter $\delta$ (defined precisely below) that measures how far the distribution is from
the critical point.

In this paper we calculate the horizontal and vertical two
point correlations of the load on the beads by generalizing the methods of ref \cite{marta}.
The horizontal load correlation 
\begin{equation}
c (m, t) = \langle w_n (t) w_{n+2m} (t) \rangle
\label{eq:horizontaldefn}
\end{equation}
is the covariance of the load on two beads that are in the same layer at depth $t$ and have a
horizontal separation $2m$. The brackets $\langle \ldots \rangle$ denote an average over the 
random fractions. By translational invariance the co-variance depends only on the separation
of the two beads ($2 m$) and not on the absolute location of either bead ($n$ and $n+2m$ respectively). 
By analogy to thermodynamic critical phenomena we would expect
the horizontal correlation length to diverge as $\xi_{{\rm horizontal}} \propto 1/\delta^\nu$ and the 
vertical correlation length to diverge as $\xi_{{\rm vertical}} \propto 1/\delta^{\nu z}$ close to the 
critical point. Furthermore we would expect the horizontal correlation function to have the universal scaling
form
\begin{equation}
c_h (m, t) - 1 = \frac{1}{\delta^\alpha} {\cal G} ( m \delta^\nu , t \delta^{\nu z}).
\label{eq:horizontalhypothesis}
\end{equation}
We find that in two dimensions the correlation function does indeed have a universal scaling form 
for large depth $t$ and small $\delta$ that we determine. Moreover we find 
the correlation length exponent $\nu = 1$, 
the dynamical exponent $z = 2$ and the amplitude exponent $\alpha = 0$. 

The vertical two point correlation 
\begin{equation}
c_v (m, t, \tau) = \langle w_n (t) w_{n + 2 m} (t + 2 \tau) \rangle
\label{eq:verticaldefn}
\end{equation}
is the covariance of the load on two beads that are located in horizontal layers separated by a depth of 
$2 \tau$ and that have a horizontal separation of $2 m$. Since the q-model is only translationally 
invariant in the horizontal direction the vertical two point correlation depends on both the separation of
the two layers ($2 \tau$) as well as the absolute location of the first layer ($t$). 
Again by analogy to thermodynamic critical phenomena we would expect the vertical correlation to have
the universal scaling form
\begin{equation}
c_v (m, t, \tau) - 1= \frac{1}{\delta^\beta} {\cal H} (m \delta^\nu, t \delta^{\nu z}, \tau \delta^{\nu z}).
\label{eq:verticalscaling}
\end{equation}
We find that in two dimensions the correlation function does indeed have a universal scaling form for large depth
$t$ and small $\delta$ that we determine. Moreover we find
the correlation length exponent $\nu=1$ and the dynamical exponent $z=2$ in agreement with the values
obtained from the horizontal correlation calculation and the amplitude exponent $\beta = 1$. 

The relationship of the q-model to random walk models can be understood as follows. 
Suppose that instead of a uniform load, a unit load is applied to a small number of beads 
in the top layer. In the singular case the subsequent propagation of the load follows the
trajectories of random walkers that coalesce upon contact and move together thereafter.
The non-singular case can be interpreted as walkers that have a certain probability to fission.
By contrast in common random walk models the walkers either pass through each other
in the non-interacting case or bounce off each other or annihilate each other upon contact in 
interacting models \cite{fisher}. It will be seen below that the horizontal correlation functions
of the q-model decay as Gaussians rather than exponentials reflecting their close
kinship to random walks. 

In the prior work of Lewandowska {\em et al.} \cite{marta} 
the critical point was characterized by computing 
the covariance of the load on a single bead. Ref \cite{marta} derived the universal scaling
function that describes the evolution of this quantity and computed a number of critical exponents
including the product $\nu z = 2$ (but not the separate values of $\nu$ and $z$). 
The present work adds to this body of knowledge the universal scaling forms for the
two point correlations and the important critical exponents $\nu$ and $z$. The present work is limited to two 
dimensions where the critical behavior is most interesting. Ref \cite{marta} also studied higher 
dimensions and thereby discovered that the upper critical dimension for the q-model is three. 
Moreover ref \cite{marta} considered the effect of an ``injection term'' that takes into account the
weight of the beads which has been neglected in comparison to the applied load in eq (\ref{eq:weightrecursion}).
In other related work Rajesh and Majumdar \cite{rajesh} computed the two point correlations for
the q-model but at large depth and far from the critical regime. Finally Snoeijer and van Leeuwen
refs \cite{snoeijer, snoeijer2} have studied the distribution of load at asymptotically large depths, as discussed
further in the conclusion.

\section{Horizontal Correlations}

\subsection{Exact Solution}

Our purpose in this subsection is to evaluate the evolution of the correlations eq (\ref{eq:horizontaldefn}) 
with depth by use of eq (\ref{eq:weightrecursion}). We begin by noting that the random fractions that
appear in eq (\ref{eq:weightrecursion}) are assumed to be independent identically distributed random 
variables with a symmetric distribution $P(f)$ that satisfies the requirement $P(f) = P(1-f)$. The 
distribution of fractions may be characterized by a parameter $\epsilon$ defined via
\begin{equation}
\langle \left(f - \frac{1}{2} \right)^2 \rangle = \frac{\epsilon}{4}.
\label{eq:epsilondefn}
\end{equation}
It is easy to verify that $\epsilon = 1$ for the singular distribution and $\epsilon = \frac{1}{3}$ for the
uniform distribution. Thus $\delta$ defined as
\begin{equation}
\delta = 1 - \epsilon
\label{eq:deltadefnd}
\end{equation}
is a measure of the distance of a distribution from the singular distribution.

We assume that a uniform load is applied to the top layer. Hence the correlation in the
top layer $c_m(m, 0) = 1$. As shown in section II A of ref \cite{marta}, 
the subsequent evolution of the correlation is governed by
\begin{eqnarray}
c_h (m, t+1) & = & \sum_n H_{mn} c (n, t) \nonumber \\
& = & \left( \frac{1}{2} + \frac{\epsilon}{2} \delta_{m,0} \right) c_h (m, t) \nonumber \\
& + & \left( \frac{1}{4} - \frac{\epsilon}{4} \delta_{m,1} \right) c_h (m-1, t) \nonumber \\
& + &\left( \frac{1}{4} - \frac{\epsilon}{4} \delta_{m,-1} \right) c_h (m+1, t).
\label{eq:evolution}
\end{eqnarray}
The evolution matrix $H$ may be interpreted as the Hamiltonian of a quantum particle on a
lattice with a non-Hermitian barrier at the origin. It was shown in section II B of ref \cite{marta} that one can
obtain an integral expression for the correlation at depth $t$ by making a bi-orthogonal expansion
in the left and right eigenstates of the non-hermitian Hamiltonian $H$.
An important subtlety in this procedure is that one has to verify that the eigenstates of $H$ are
complete since there is in general no guarantee that a non-Hermitian operator will have a complete
set of eigenstates. Making use of the bi-orthogonal expansion and the expressions for the left and
right eigenvectors of $H$ derived in ref \cite{marta} we obtain 
\begin{eqnarray}
c_h (m, t) & = & 1 + \int_{-\pi}^{\pi} \frac{d k}{2 \pi} \; \cos^{2t} \left( \frac{k}{2} \right) e^{i k m}
\times \nonumber \\
& & \frac{ \left[ - i \epsilon^2 \sin k - \epsilon (1 - \epsilon) (1 + \cos k) \right] }{\left[ (1 - 2 \epsilon + 2 \epsilon^2) 
+ (1 - 2 \epsilon) \cos k \right]}
\label{eq:horizontalexact}
\end{eqnarray}
Eq (\ref{eq:horizontalexact}) is an exact expression for the horizontal correlation function and is the
main result of this subsection. We note that eq (\ref{eq:horizontalexact}) applies for all $m$ except $m=0$.
For $m=0$ the horizontal correlation function $c_h (0, t)$ coincides with the variance calculated by ref
\cite{marta} and the expression for $c_h (0, t)$ is given by their eq (41).

\subsection{Critical Exponents and Scaling Limit}

To obtain the scaling limit of the correlation function 
we now simplify eq (\ref{eq:horizontalexact}), assuming $t \gg 1$ and $\delta \ll 1$, but without making
any assumptions about the relative size of $t$, $\delta$ and $m$\footnote{First, approximate
$\cos^{2t} (k/2) \approx \exp( - k^2 t/4)$. This form is justified for small $k$ and leads to negligible
error for large $k$ since both the exact expression and the approximation are negligible for $k$
large compared to $1/\sqrt{t}$.  For the same reason we may extend the range of integration in 
eq (\ref{eq:horizontalexact}) to infinity with negligible error. Finally since the integrand has negligible
weight for large $k$ we may expand both numerator and denominator in the second line of eq (\ref{eq:horizontalexact})
to leading order in $k$ and $\delta$. Eqs (\ref{eq:horizontalscaling})
and (\ref{eq:fdefnd}) result after making these approximations
and a rescaling of the integration variable.}. We find 
\begin{equation}
c_h(m, t) - 1 = - f(\theta, \mu) - \frac{\partial}{\partial \mu} f( \theta, \mu)) 
\label{eq:horizontalscaling}
\end{equation}
where $\theta = \delta \sqrt{t}$ is a scaled measure of depth and $\mu = 2 m \delta$ is a scaled
measure of horizontal separation and the function 
\begin{equation}
f(\theta, \mu) = \int_{-\infty}^{\infty} \frac{d u}{\pi} \; \frac{e^{-u^2 \theta^2} e^{- i \mu u} }{1 + u^2}.
\label{eq:fdefnd}
\end{equation}
It is evident from eqs (\ref{eq:horizontalscaling}) and (\ref{eq:fdefnd}) that the horizontal correlations 
have the expected scaling form eq (\ref{eq:horizontalhypothesis}). Furthermore, from the form of the
scaled variables $\theta$ and $\mu$ we infer that the horizontal correlation length 
exponents $\nu = 1$ and the dynamical exponent $z = 2$. 

To derive the asymptotic behavior of the correlation function in the scaling regime it is convenient 
to rewrite the correlation function as
\begin{equation}
c_h (m, t) - 1 = - \frac{1}{\sqrt{\pi}} \int_{\mu/\theta}^\infty d s \exp \left[ - \theta \left( s - \frac{\mu}{\theta} 
\right) \right] \exp \left( - \frac{1}{4} s^2 \right). 
\label{eq:alternatehorizontal}
\end{equation}
This form is obtained by noting that $f(\theta, \mu)$ is an integral over a product of a Gaussian and a
Lorentzian. Using Parseval's theorem it may therefore be written as an integral over a product of the
Fourier transforms of the Gaussian and Lorentzian factors. Using this transformed
representation for $f(\theta, \mu)$ leads from eq (\ref{eq:horizontalscaling}) to eq (\ref{eq:alternatehorizontal}).

Now let us consider the regime of small depth, $\theta \ll 1$ or equivalently $t \delta^2 \ll 1$
(the ``critical regime''). 
In this regime we would expect the correlation function to be indistinguishable from the
correlation function at the critical point. Indeed we find that for small $\theta$ eq (\ref{eq:alternatehorizontal})
simplifies to 
\begin{equation}
c_h (m, t) - 1 = - \frac{1}{\sqrt{\pi}} \int_{\mu/\theta}^{\infty} d s \exp \left( - \frac{1}{4} s^2 \right).
\label{eq:shallowhorizontal}
\end{equation}
Thus the correlation function depends only on $\mu/\theta$ or equivalently $m/\sqrt{t}$ and is 
independent of $\delta$. Eq (\ref{eq:shallowhorizontal}) reveals that in the critical regime the
loads (or to be precise the deviations in load from the mean) on neighboring beads are strongly
{\em anti-correlated}. For small distances
($m/\sqrt{t}$) eq (\ref{eq:shallowhorizontal}) simplifies to
\begin{equation}
c_h (m, t) - 1 \approx -1 +  \frac{2 m}{\sqrt{\pi t}} + \ldots
\label{eq:shallowhorizontalshort}
\end{equation}
At large distances ($m/\sqrt{t} \gg 1$) eq (\ref{eq:shallowhorizontal}) simplifies to
\begin{equation}
c_h (m,t) - 1 \approx - \sqrt{ \frac{t}{\pi m^2} } \exp \left( - \frac{ m^2 }{t} \right);
\label{eq:shallowhorizontallong}
\end{equation}
in other words the anti-correlations decay as a Gaussian. 
In summary we find that in the critical regime there are strong horizontal anti-correlations
whose range grows with the square root of the depth.

Next let us consider the regime of large depth, $\theta \gg 1$ or equivalently $t \delta^2 \gg 1$
(the ``saturated regime''). In this regime eq (\ref{eq:alternatehorizontal}) simplifies to 
\begin{equation}
c_h (m, t) = - \frac{ 1}{\sqrt{ \pi t \delta^2 } } \exp \left( - \frac{m^2}{t} \right).
\label{eq:deephorizontal}
\end{equation}
Thus in the saturated regime there are weak anti-correlations. The amplitude of these
anti-correlations falls off inversely with depth whilst their range continues to grow as the square 
root of the depth.

\section{Vertical Correlations}

\subsection{Exact Solution}

There is a remarkably simple exact relationship between the horizontal and vertical 
correlations for the q-model \cite{rajesh}. In order to establish this relationship it is
instructive to first consider the correlations
between beads in two consecutive layers, $t$ and $t+1$, with a horizontal separation $2m+1$. By
use of the evolution eq (\ref{eq:weightrecursion}) we may write
\begin{eqnarray}
\langle w_n (t) w_{n+2m+1}(t+1) \rangle & = & 
\langle f_{n+2m+2,t} w_n (t) w_{n+2m+2}(t) \rangle \nonumber \\
&+ &
\langle (1 - f_{n+2m, t} ) w_{n} (t) w_{n + 2m} (t) \rangle \nonumber \\
\label{eq:onelayer}
\end{eqnarray}
Note that the load on the beads in layer $t$ is independent of the fractions
that appear on the right hand side of eq (\ref{eq:onelayer}); these fractions
determine the subsequent propagation of load from layer $t$ propagates to layer $t+1$. 
Hence the averages on the right hand side of eq (\ref{eq:onelayer}) factorize. Recalling
that by virtue of the symmetry condition on the distribution of the fractions,
$\langle f \rangle = \langle (1 - f) \rangle = 1/2$, and making use of the definition of the
horizontal correlation eq (\ref{eq:horizontaldefn}) 
we obtain
\begin{equation}
\langle w_n (t) w_{n+2m+1}(t+1) \rangle = 
\frac{1}{2} c_h (2m + 2, t) + \frac{1}{2} c_h (2m, t).
\label{eq:onefactorized}
\end{equation}
Similarly one can relate the correlation between beads separated by two layers,
$ c_v (m, t, \tau \rightarrow 1)$, to horizontal correlations 
by twice using the evolution eq (\ref{eq:weightrecursion}) to obtain
\begin{equation}
c_v (m, t, 1) = \frac{1}{4} c_h (m + 2, t) + \frac{1}{2} c_h (m, t) + \frac{1}{4} c_h (m - 2, t). 
\label{eq:twofactorized}
\end{equation}
By now the astute reader will have noted the appearance of binomial coefficients in 
eqs (\ref{eq:onefactorized}) and (\ref{eq:twofactorized}) and indeed it is not difficult to
show by induction that
\begin{equation}
c_v (m, t, \tau) = \frac{1}{2^{2 \tau}} \sum_{k = - \tau}^{+\tau} 
\left( \begin{array}{c}
2 \tau \\
\tau + k 
\end{array}
\right) c_h (m - k, t).
\label{eq:verticalexact}
\end{equation}
Eq (\ref{eq:verticalexact}) is the exact relationship between vertical and horizontal 
correlations and is the main result of this subsection.

\subsection{Critical Exponents and Scaling Limit}

In this section we derive the universal scaling behavior of the vertical correlation function. 
There are three circumstances to consider. It is clear from fig \ref{fig:qmodel} that each bead
sits at the bottom of an inverted cone of beads that are partially supported by it. The correlation
between two beads will evidently be strong if the upper bead of the pair lies inside the support 
cone of the lower bead. Under this circumstance, which we call the ``direct case'', $|m| \leq \tau$. 
On the other hand if the two beads are so far apart that their support cones do not intersect at all
the correlation between their loads is rigorously zero. The ``uncorrelated case'' corresponds to 
the condition $|m| > 2 t + \tau$. Finally there is the intermediate ``indirect case'' in which the two
support cones do intersect but the upper bead does not lie inside the support cone of the lower bead.

We first consider the scaling limit for the direct case. The sum in eq (\ref{eq:verticalexact}) can be 
separated into two parts. The first part corresponds to the single term with $k = m$. This term
determines the contribution of the covariance in load of the upper bead to the vertical correlation
with the lower bead.
It will emerge that this co-variance contribution leads to positive correlations and tends to dominate the
direct vertical correlations. The second part corresponds to the remaining terms in eq (\ref{eq:verticalexact}).
It will emerge that this contribution leads to anti-correlations and is generally subdominant. 

We now compute the co-variance contribution to the direct vertical correlation. Retaining only the
$k = m$ term in eq (\ref{eq:verticalexact}) we obtain
\begin{equation}
c_v (m, t, \tau) = \frac{1}{2^{2 \tau}} \left( 
\begin{array}{c}
2 \tau \\
\tau + m 
\end{array}
\right) c_h (0, t).
\label{eq:verticalcone}
\end{equation}
Assuming that $\tau \gg 1 $ and $|m| \gg 1$ we may approximate the binomial coefficient
as a Gaussian. Finally using the scaling limit of $c_h (0,t)$ given in eq (48) of ref \cite{marta} 
we obtain \footnote{The alert reader will note that we have also transformed eq (48) of ref \cite{marta},
which features an integral of the product of a Gaussian and a Lorentzian, 
into an integral over the corresponding Fourier transforms instead
by use of Parseval's theorem.}
\begin{eqnarray}
c_v (m, t, \tau) - 1& = & 
\frac{1}{\sqrt{\pi \tau \delta^2}} \exp \left[- \frac{(m \delta)^2}{\tau \delta^2} \right] \nonumber \\
& \times &
\left[ 1 - \frac{1}{\sqrt{\pi}} \int_0^\infty d s \exp( - s \sqrt{t \delta^2} ) 
e^{ - s^2/4 } \right].
\nonumber \\
\label{eq:directverticalscaling}
\end{eqnarray}
Eq (\ref{eq:directverticalscaling}) gives the covariance contribution to the 
the direct vertical correlations in the scaling limit.  
It agrees with the conjectured universal scaling form in eq (\ref{eq:verticalscaling}) and we find that the 
critical exponents are $\beta = 0, \nu = 1, z = 2$. 

To gain further insight into the covariance contribution to the 
direct vertical correlations it is useful to consider the limiting behavior 
of eq (\ref{eq:directverticalscaling}) in the
critical regime ($t \delta^2 \ll 1$) and the saturated regime ($t \delta^2 \gg 1$). In the former limit 
the dependence on $\delta$ cancels and we find
\begin{equation}
c_v (m, t, \tau) - 1 = \frac{2}{\pi} \sqrt{ \frac{t}{\tau} }
\exp \left[ - \frac{ m^2 }{ \tau } \right]
\label{eq:verticalshallowscaling}
\end{equation}
Thus in the critical regime there are strong vertical correlations (not anti-correlations) and these correlations
grow with the square root of the depth $t$. 
The horizontal range of the correlations grows as the square root of the
vertical separation between the two beads, $\tau$. If we imagine increasing the
horizontal separation of the beads, keeping the vertical separation fixed, then the correlations die away well 
before we reach the boundary of the support cone. In the saturated regime we find 
\begin{equation}
c_v (m, t, \tau) = \frac{1}{\delta} \frac{1}{\sqrt{\pi \tau} } \exp \left( - \frac{m^2}{\tau} \right).
\label{eq:verticaldeepscaling}
\end{equation}
Here too there are strong vertical correlations (not anti-correlations) but these correlations have
saturated and are no longer growing with depth $t$. Their horizontal range grows as the square root
of the vertical separation between the beads, $\tau$. The correlations die away well before we
reach the boundary of the support cone, just as they do in the critical regime. 

We turn now to the second contribution to the vertical correlations in the direct case,
namely the sum over terms with $k \neq m$ in eq (\ref{eq:verticalexact}). 
In the scaling regime we may approximate the
binomial coefficients in eq (\ref{eq:verticalexact}) as a Gaussian, convert the sum into an
integral and extend the range of integration to infinity to obtain
\begin{equation}
c_v (m, t, \tau) = \int_{-\infty}^{\infty} d k \; \frac{1}{\sqrt{\pi \tau}} \exp \left( - \frac{k^2}{\tau} \right)
c_h (m - k, t). 
\label{eq:intermediatestep}
\end{equation}
Using eq (\ref{eq:alternatehorizontal}) for $c_h$ we may perform the integral over $k$ explicitly to obtain
\begin{eqnarray}
c_v (m, t, \tau) - 1&  = &  - \frac{1}{\sqrt{ \pi}} \exp \left[ - \frac{ (m \delta )^2 }{ (t + \tau) \delta^2 } \right]
\int_0^\infty 
d s \; e^{ - s^2/4 } \nonumber \\
& \times & 
\exp \left[ - s \left( 
\sqrt{ (t + \tau) \delta^2 } 
- \frac{ m \delta }{ \sqrt{ (t + \tau) \delta^2 } } 
\right) \right].
\nonumber \\
\label{eq:indirectscaling}
\end{eqnarray}
Note that this contribution to the vertical correlation function is also of the conjectured form eq (\ref{eq:verticalscaling})
with the same critical exponents $\beta = 0, \nu = 1$ and $z=2$. 
It is noteworthy that this contribution is a function only of the $(t + \tau) \delta^2$ and the ratio $m \delta/\sqrt{(t + 
\tau) \delta^2}$; it is a more restricted 
function of its arguments than required by the scaling form eq (\ref{eq:verticalscaling}).
The full vertical correlation in the direct case is the 
sum of eq (\ref{eq:directverticalscaling}) and eq (\ref{eq:indirectscaling}); however, as noted above and shown below,
the contribution of eq (\ref{eq:indirectscaling}) is generally dominated by the contribution of eq 
(\ref{eq:directverticalscaling}).

To gain further insight into the second contribution to the direct vertical correlations it is useful to consider
the limiting behavior of eq (\ref{eq:indirectscaling}) in the critical regime ($t \delta^2 \ll 1$) and the saturated
regime ($t \delta^2 \gg 1$). In the former limit the dependence on $\delta$ cancels and we find
\begin{eqnarray}
c_v (m, t, \tau) - 1 & = &  - \frac{1}{\pi} \exp \left( - \frac{m^2}{t + \tau} \right) 
\int_0^\infty d s \; e^{-s^2/4} \nonumber \\
& & \times \exp \left( \frac{ m s }{ \sqrt{ t + \tau} } \right).
\label{eq:indirectscalingcritical}
\end{eqnarray}
Eq (\ref{eq:indirectscalingcritical}) is still a formidable expression but it simplifies to 
$c_v (m, t \tau) - 1 \approx - 1$ for $m/\sqrt{t+\tau} \ll 1$ and to $c_v (m, t, \tau) - 1 \approx - 2$ for
$m/\sqrt{t + \tau} \gg 1$. Thus the second contribution to the direct vertical correlation in the critical
regime corresponds to a weak anti-correlation that is negligible compared to the strong positive
correlation implied by the first contribution, eq (\ref{eq:verticalshallowscaling}).
Next let us consider the saturated regime ($t \delta^2 \gg 1$). In this regime eq (\ref{eq:indirectscaling})
simplifies to 
\begin{equation}
c_v (m, t, \tau) - 1 = - \frac{1}{\delta} \frac{1}{\sqrt{ \pi (t + \tau) } } \exp \left( - \frac{ m^2 }{ t + \tau } \right).
\label{eq:indirectscalingsaturated}
\end{equation}
Eq (\ref{eq:indirectscalingsaturated}) should be compared to the first contribution eq (\ref{eq:verticaldeepscaling}).
Thus in the saturated regime too the second contribution to the direct vertical correlations is smaller
than the first contribution. 

In summary, in the direct case the vertical correlation is the sum of two contributions both of which have the
conjectured universal scaling form, eq (\ref{eq:verticalscaling}), 
with exponents $\beta = 0, \nu = 1$ and $z = 2$. The first contribution
eq (\ref{eq:directverticalscaling}) 
corresponds to a positive correlation and generally dominates. The second contribution eq (\ref{eq:indirectscaling})
corresponds to an anti-correlation and is generally subdominant. It is illuminating to look at the limiting
behavior of the two components in the critical regime ($t \delta^2 \ll 1$) and the saturated regime.
In the critical regime the first contribution is given by 
eq (\ref{eq:verticalshallowscaling}) and the second contribution, which lies between
$-1$ and $-2$, is given by eq (\ref{eq:indirectscalingcritical}). In the saturated regime the first contribution is given by 
eq (\ref{eq:verticaldeepscaling}) and the 
second contribution by eq (\ref{eq:indirectscalingsaturated}).

Finally, we now turn to the vertical correlation in the indirect case, $|m| > \tau$. 
In this case the first contribution is absent since the
term $k = m$ lies outside the range of the sum in eq (\ref{eq:verticalexact}).
Thus in the indirect case there is only a weak anti-correlation between beads 
given by eq (\ref{eq:indirectscaling}). This concludes our analysis of the vertical correlations.

\section{Conclusion}

In this paper we add to the existing body of known results about the q-model
by obtaining new results on the exact horizontal and vertical correlation functions. We 
show that the correlation functions have a universal scaling form close to the critical 
point identified in previous work \cite{marta} and we determine the scaling form and  
associated critical exponents. 
It is intriguing that the q-model, which describes physics far from equilibrium, nonetheless
shows behavior reminiscent of equilibrium thermodynamic critical phenomena. 
Among the many questions that our work leaves open we here mention two.
First we note that it may be possible to calculate three point and higher correlation
functions by means of a non-hermitian generalization of Bethe ansatz \cite{me}. Second,
it is desirable to determine the dynamics of the entire distribution of
load on a single bead. Ref \cite{marta} obtained the dynamics of the entire distribution
right at the critical point and Snoeijer and van Leewen \cite{snoeijer2} studied the large depth asymptotic dynamics 
of the load distribution in a case far from the  the critical point. Close to the critical point 
ref \cite{marta} made a scaling hypothesis about the form of this distribution and it 
remains of interest to verify this conjecture via numerics or exact solution.

Alex St. John acknowledges support by the REU program at Case Western Reserve University
via NSF grant DMR-0850037. Harsh Mathur acknowledges support by the DOE. We also acknowledge
valuable past discussions with Onuttom Narayan that motivated the present work and wish to thank
Philip Taylor for discussion of this work and comments on the manuscript.


\begin{thebibliography}{99}
\bibitem{coppersmith} S.N. Coppersmith, C.-H. Liu, S.N. Majumdar, O. Narayan and T.A. Witten, Phys Rev {\bf E53},
4673 (1996).
\bibitem{takayasu} H. Takayasu, Phys Rev Lett {\bf 63}, 2563 (1989); H. Takayasu, I. Nishikawa, and H. Tasaki,
Phys Rev {\bf A37}, 3110 (1988). 
\bibitem{chalker} J.T. Chalker and A. Dohmen, Phys Rev Lett {\bf 75}, 4496 (1995). 
\bibitem{siggia} For a review, see B.I. Shraiman and E.D. Siggia, Nature {\bf 405}, 639 (2000). 
\bibitem{dodds} A.E. Scheidegger, Bull. Intl. Assoc. Sci. Hydrol {\bf 12}, 15 (1967);
P.S. Dodds and D.H. Rothman, Phys Rev {\bf E59}, 4865 (1999).
\bibitem{marta} M. Lewandowska, H. Mathur and Y.K. Yu, Phys Rev {\bf E64}, 026107 (2001). 
\bibitem{experiments} C.-H. Liu, S.R. Nagel, D.A. Schecter, S.N. Coppersmith, S.N. Majumdar, 
O. Narayan, and T.A. Witten, Science {\bf 269}, 513 (1995); D.M. Mueth, H.M. Jaeger, and S.R. Nagel, 
Phys Rev {\bf E57}, 3164 (1998).
\bibitem{fisher} M.E. Fisher, J Stat Phys {\bf 34}, 667 (1984). 
\bibitem{rajesh} R. Rajesh and S.N. Majumdar, Phys Rev {\bf E62}, 3186 (2000).
\bibitem{snoeijer} J.H. Snoeijer, J.M.J. van Leeuwen, Phys Rev {\bf E65}, 051306 (2002). 
\bibitem{snoeijer2} J. H. Snoeijer and J.M.J. van Leeuwen, J Stat Phys {\bf 109}, 449 (2002).
\bibitem{me} H. Mathur, unpublished.

\end{thebibliography}
\end{document}